\documentclass[preprint,12pt]{elsarticle}

\usepackage{amssymb}
\usepackage{amsthm}

\usepackage{lineno}

\usepackage{amsmath}
\usepackage{color}
\usepackage{subcaption}

\usepackage[colorlinks=false, hidelinks]{hyperref}
\usepackage{float} 
\usepackage{booktabs} 

\usepackage[top=2.5cm,bottom=2.5cm,left=3cm,right=3cm]{geometry}

\journal{Statistics in Medicine (pre-peer review; manuscript was accepted).}

\begin{document}
\begin{frontmatter}



\title{A studentized permutation test for three-arm trials in the `gold standard' design}


\author[add1]{Tobias M\"utze\corref{cor1}}
\author[add1,add2]{Frank Konietschke}
\author[add3,add4]{Axel Munk}
\author[add1]{Tim Friede}

\address[add1]{Department of Medical Statistics, University Medical Center G{\"o}ttingen, G{\"o}ttingen, Germany}
\address[add2]{Department of Mathematical Sciences, University of Texas at Dallas, Richardson, TX, USA}
\address[add3]{Institute for Mathematical Stochastics, Georg-August-University of G{\"o}ttingen, G{\"o}ttingen, Germany}
\address[add4]{Max Planck Institute for Biophysical Chemistry, G{\"o}ttingen, Germany}

%
 
\cortext[cor1]{Correspondence: tobias.muetze@med.uni-goettingen.de}


\begin{abstract}
The `gold standard' design for three-arm trials refers to trials with an active control and a placebo control in addition to the experimental treatment group. 
This trial design is recommended when being ethically justifiable and it allows the simultaneous comparison of experimental treatment, active control, and placebo.
Parametric testing methods have been studied plentifully over the past years.
However, these methods often tend to be liberal or conservative when distributional assumptions are not met particularly with small sample sizes. 
In this article, we introduce a studentized permutation test for testing non-inferiority and superiority of the experimental treatment compared to the active control in three-arm trials in the `gold standard' design. 
The performance of the studentized permutation test for finite sample sizes is assessed in a Monte-Carlo simulation study under various parameter constellations. 
Emphasis is put on whether the studentized permutation test meets the target significance level. 
For comparison purposes, commonly used Wald-type tests are included in the simulation study. 
The simulation study shows that the presented studentized permutation test for assessing non-inferiority in three-arm trials in the `gold standard' design outperforms its competitors for count data.
The methods discussed in this paper are implemented in the R package {\tt{ThreeArmedTrials}} which is available on the comprehensive R archive network (CRAN).
\end{abstract}


\end{frontmatter}



\section{Introduction}
Clinical trials with the gold standard design include an active reference and a placebo in addition to an experimental treatment. 
In various areas in clinical research this design became increasingly popular \cite{melosky2015pan, bossche2015multi, bines2015safety, hoopfer2015three}.
Of particular interest in trials with the gold-standard design is assessing non-inferiority or superiority of the experimental treatment compared to the reference treatment. 
With $\mu_{E}$, $\mu_{R}$, and $\mu_{P}$ the parameter of interest for the experimental treatment group, the reference treatment group, and the placebo group, respectively, and smaller parameters $\mu_{k}$ being desired, the null hypothesis of non-inferiority and superiority can be defined by 
\begin{align*}
H_0: \mu_{P}-\mu_{E} \leq \Delta(\mu_{P}-\mu_{R} ) 
\quad \text{vs.} \quad
H_1: \mu_{P}-\mu_{E} > \Delta(\mu_{P}-\mu_{R} ).
\end{align*}
Superiority is tested with a margin of $\Delta\in [1,\infty)$ and non-inferiority with a margin of $\Delta \in (0,1)$. 
In case of testing non-inferiority, the hypothesis above is commonly referred to as the \textit{retention of effect hypothesis} \cite{mielke2008assessment}. 
This definition of non-inferiority and superiority assumes that the parameter of the reference group $\mu_{R}$ is larger than the parameter of the placebo group $\mu_{P}$, that is $\mu_{R}>\mu_{P}$. 
This assumption can also be interpreted as assay sensitivity, that is the ability of a clinical trial to distinguish an effective from an ineffective treatment \cite{ichE10}.
In general, testing assay sensitivity might be required in addition the assessment of non-inferiority or superiority. 
In this manuscript, however, we focus on testing the hypothesis $H_0$ for either non-inferiority or superiority.
Testing strategies in the gold standard design have been studied in detail by Koch and R\"ohmel \cite{koch2004hypothesis}.
For a detailed discussion of the usage of placebo in clinical trials, we refer to already existing literature \cite{decHel, hill1994, temple2000}.\\ \indent
The retention of effect hypothesis, that is the hypothesis $H_0$ in the case of non-inferiority, has already been studied for various endpoint scales. 
Pigeot \textit{et al.} \cite{pigeot2003assessing},  Hauschke and Pigeot  \cite{hauschke2005establishing}, and Hasler \textit{et al.} \cite{hasler2008assessing} studied the retention of effect hypothesis for normally distributed endpoints. 
Additionally, the retention of effect hypothesis has been studied by Kieser and Friede  \cite{kieser2007planning} and by Munk \textit{et al.} \cite{munk2007testing} for binary data, Mielke \textit{et al.} \cite{mielke2008assessment} investigated censored exponentially distributed endpoints,  and M\"utze \textit{et al.}  \cite{muetze2015design} for negative binomially distributed responses.
Moreover, Mielke and Munk  \cite{mielke2009assessment} and Balabdaoui \textit{et al.}  \cite{balabdaoui2009} established the maximum-likelihood theory  for parametric models and a generalization of the retention of effect hypothesis. 
Munzel  \cite{munzel2009nonparametric} derived a statistical test for the non-parametric equivalent. 
Kombrink \textit{et al.} \cite{kombrink2013design} introduced a semiparametric analysis of three-arm trials for censored time-to-event data. 
A Bayesian approach to three-arm non-inferiority trials has been proposed by Ghosh \textit{et al.} \cite{ghosh2011assessing}.\\ \indent
The distributional assumptions being made for the listed methodology for testing the retention of effect hypothesis $H_0$ cannot always be assessed, for example when studies are small or when overdispersion is present. 
If the assumptions for a parametric test are not fulfilled, the test is in general either conservative or liberal. 
Furthermore, likelihood based tests -- such as Wald-type or score tests -- are in general neither robust nor efficient when the model is misspecified.
These limitations of likelihood based tests can be bypassed by relying on non-parametric tests (see for example Freitag \textit{et al.} \cite{freitag2006non}). One class of non-parametric tests are studentized permutation tests.
Studentized permutation tests are exact tests if the random variables are exchangeable and if exchangeability is not given they control the significance level asymptotically in many settings. 
Studentized permutation tests do not require the assumption of any specific distribution resulting in widely applicable and robust tests.
Moreover, even though they are asymptotic tests, studentized permutation tests are often advantageous for small sample sizes compared to Wald-type approaches.
The general theory of studentized permutation tests has been studied for non-\textit{i.i.d.} random variables by Janssen  \cite{janssen1997studentized}.
Studentized permutation tests have already been studied for the generalized Behrens-Fischer problem by Janssen  \cite{janssen1997studentized}, the non-parametric Behrens-Fisher problem with unpaired data by Neubert and Brunner  \cite{neubert2007studentized} and with paired data by Konietschke and Pauly  \cite{konietschke2012studentized}, for spatial point patterns by Hahn  \cite{hahn2012studentized}, for randomly right censored data by Brendel \textit{et al.} \cite{brendel2014weighted}, for heteroscedastic two-sample problems by Janssen and Pauls \cite{janssen2005monte}, and for general factorial designs by Pauly \textit{et al.}  \cite{pauly2015asymptotic}.
The aim of this manuscript is to establish a permutation test for non-inferiority and superiority in three-arm trials which does not require any specific distributional assumption and performs well for small sample sizes. \\ \indent
This article is structured as follows. 
In Section 2, an example from clinical trials in multiple sclerosis is discussed.
The statistical model and the theory of the studentized permutation test for the retention of effect hypothesis will be introduced in Section 3. 
The finite sample size properties of the studentized permutation test for the hypothesis $H_0$ are studied by means of simulation studies in Section 4. 
The article concludes with a discussion of our findings in Section 5 and proofs of the asymptotic properties of the studentized permutation test in the appendix. 
\section{A clinical trial example}
In a clinical trial setting, the hypothesis $H_0$ has been studied for endpoints in trials with asthmatic patients where the considered trial consists of 74 subjects \cite{pigeot2003assessing}. 
Other examples include clinical trials in depression \cite{kieser2007planning, mielke2009assessment, kombrink2013design} and multiple sclerosis \cite{muetze2015design}.
The theoretical work about the studentized permutation test in this publication, however, is motivated by phase II clinical trials in relapsing-remitting multiple sclerosis.  
Multiple sclerosis is an inflammatory disease of the brain and spinal cord \cite{MSlancet}. 
Relapsing-remitting multiple sclerosis is a form of multiple sclerosis which is characterized by sudden worsenings of the symptoms called relapses.
Clinical trials, and particularly phase II studies, in relapsing-remitting multiple sclerosis generally include multiple treatment arms as well as placebo arms and can be rather small \cite{sorensen2014safety}.
The number of subjects per group can be as small as ten subjects \cite{sorensen2014safety, pakdaman2006treatment, steinvorth2013explaining, llufriu2014randomized, rover2015changing}. 
A phase II trial with active and placebo control in relapsing-remitting multiple sclerosis has be published by Kappos \textit{et al.}  \cite{kappos2011ocrelizumab}. 
Besides the active control arm (intramuscular interferon beta-1a) and the placebo, the trial includes two arms with different doses of the monoclonal antibody ocrelizumab. 
The trial aims to assess the safety and efficacy of ocrelizumab. 
The primary endpoint was the total number of gadolinium-enhancing T1 lesions.
Lesions are the regions in the brain and spinal cord which have been damaged by the disease.
The results of  the total number of gadolinium-enhancing T1 lesions are summarized in Table \ref{Table:ResultsKappos}.
\begin{table}[ht]
\begin{center}
\caption{Total number of galodinium-enhancing T1 lesions over weeks  12, 16, 20, and 24 \cite[Table 2]{kappos2011ocrelizumab}.}
\label{Table:ResultsKappos}
\scriptsize
\begin{tabular}{l*{1} {c}{c}{c}{c}}
\toprule
\textbf{Number of lesions}& \textbf{Placebo} &  \textbf{Ocrelizumab 600 mg} & \textbf{Ocrelizumab 2000 mg} & \textbf{Interferon beta-1a} \\ \midrule
0 & 19 $(35\%)$ & 39 $(77\%)$ & 43 $(82.7\%)$ & 25 $(49\%)$ \\
1 & 6 $(11\%)$ & 2 $(4\%)$ & 6 $(11.5\%)$ & 5 $(10\%)$ \\
2 & 7 $(13\%)$ & 6 $(6\%)$ & 1 $(1.9\%)$  & 5 $(10\%)$ \\
3 & 3 $(6\%)$ & 0 & 2 $(3.8\%)$ & 0 \\
$\geq 4$ & 19 $(35\%)$ & 4 $(8\%)$ & 0 & 17 $(33\%)$ \\ \hline
n & 54 & 51 & 52 & 52 \\
Mean & 5.5 & 0.6 & 0.2 & 6.9 \\
Standard Deviation & 12.5 & 1.5 & 0.7 & 16 \\
\bottomrule
\end{tabular}
\end{center}
\end{table}
As Table \ref{Table:ResultsKappos} highlights, the distribution of the lesion is characterized by having high probability of zeros, between $35\%$ and $82.7\%$ in the example, but also a heavy tail, for instance in the example trial four or more enhanced lesions occurred between $0\%$ and $35\%$ of all cases depending on the group. 
In literature it has been proposed to model the galodinium-enhancing T1 lesions as overdispersed count data, in particular as negative binomially distributed \cite{sormani1999modelling}.  
Overdispersion refers to the variance exceeding the mean.
In Table \ref{Table:KapposNegBin} the theoretically expected values of a negative binomial distribution with expectation and variance as in the respective sample  are compared to observed total numbers of lesions.
\begin{table}[ht]
\caption{The expected number of lesions according to a negative binomial distribution with mean and variances as in the sample compared to the observed total number of galodinium-enhancing T1 lesions which is shown in brackets.}
\label{Table:KapposNegBin}
\begin{center}
\scriptsize
\begin{tabular}{l*{1} {c}{c}{c}{c}}
\toprule
\textbf{Number of lesions}& \textbf{Placebo} &  \textbf{Ocrelizumab 600 mg} & \textbf{Ocrelizumab 2000 mg} & \textbf{Interferon beta-1a} \\ \midrule
0 & 28 (19) & 38 (39) & 46 (43) & 26 (25) \\
1 & 5 (6) & 6 (2) & 4 (6) & 5 (5) \\
2 & 3 (7) & 3 (6) & 1 (1) & 3 (5) \\
3 & 2 (3) & 2 (0) & 1 (3) & 2 (0)\\
$\geq 4$ & 16 (19) & 2 (4) & 0 (0) & 16 (17) \\ 
\bottomrule
\end{tabular}
\end{center}
\end{table}
Table \ref{Table:KapposNegBin}, however, underlines the uncertainty concerning the distribution of the total number of galodinium-enhancing T1 lesions. 
The lesion distribution for the two doses of ocrelizumab and the active control interferon beta-1a  is described well by a negative binomial distribution. 
In contrast, the lesion distribution in the case of the placebo is clearly overdispersed but cannot be approximated accurately using a negative binomial. 
A similar observation was made when comparing  the number of lesions in the placebo group to a zero-inflated Poisson distribution.
For those cases in which the distribution of the endpoint differs between groups or where the sample size is too small to accurately assess the distribution, the studentized permutation test for the hypothesis should be preferred over a parametric approach.
\section{Statistical model and the proposed studentized permutation test}
Let the independent real-valued random variables $X_{k,i}$ with $i=1,\ldots,n_{k}$ and $k=E,R,P$ model the outcomes under experimental treatment (E), reference treatment (R), and placebo (P). 
We denote the total sample size by $n=n_{E}+n_{R}+n_{P}$. 
The random variable $X_{k,i}$  follows a distribution $F_{k}$ and has finite mean $\mathbb{E}[X_{k,i}]=\mu_k$, finite positive variance  $\operatorname{Var}[X_{k,i}]=\sigma^2_k>0$, and finite forth moment $\mathbb{E}[X_{k,i}^{4}]$.
For any asymptotic consideration throughout this article, we assume that none of the groups vanishes asymptotically, that is $w_k=\lim_{n_k, n\to \infty} n_k/n \in (0,1)$. 
The non-inferiority or superiority hypothesis 
\begin{align*}
H_{0}:\mu_{P}-\mu_{E}\leq \Delta(\mu_{P}-\mu_{R}) 
\quad \text{vs.} \quad
H_{1}:\mu_{P}-\mu_{E}> \Delta(\mu_{P}-\mu_{R})
\end{align*}
origins from defining the margin $\delta>0$ in the non-inferiority/superiority testing problem 
\begin{align*}
H_{0}^{\delta}:\mu_{E} - \delta \geq  \mu_{R}
\quad \text{vs.} \quad
H_{1}^{\delta}:\mu_{E} - \delta < \mu_{R}
\end{align*}
as the difference of placebo response and the reference treatment response \cite{koch2004hypothesis}. 
More precisely, if the margin is defined as $\delta=f(\mu_{P}-\mu_{R})$ with $f\in(0,1)$, the hypothesis $H_0$ is equivalent to the testing problem 
\begin{align*}
H_{0}^{\delta}:\mu_{E}- f(\mu_{P}-\mu_{R}) \geq \mu_{R}
\quad \text{vs.} \quad
H_{1}^{\delta}:\mu_{E}- f(\mu_{P}-\mu_{R}) < \mu_{R}.
\end{align*}
Substituting $f$ by $1-\Delta$ results in the non-inferiority or superiority hypothesis $H_{0}$ defined above. 
It should be noted that this hypothesis can be rearranged to
\begin{align}
\label{Eq:H0_rearranged}
H_{0}:\mu_{E} - \Delta \mu_{R} + (\Delta-1)\mu_P \geq 0
\quad \text{vs.} \quad
H_{1}:\mu_{E} - \Delta \mu_{R} + (\Delta-1)\mu_P < 0.
\end{align}
The random vector $\mathbf{X}_n=(X_{n,i})_{i\leq n}\in \mathbb{R}^n$ is defined as
\begin{align}\label{eqn:randVec}
\mathbf{X}_n=(X_{E,1},\ldots,X_{E,n_{E}},X_{R,1},\ldots,X_{R,n_{R}},X_{P,1},\ldots,X_{P,n_{P}})^{\prime}
\end{align}
and  $\mathbb{P}$ denotes its probability measure. 
The mean of the random variables in group $k=E, R, P$ is given by $\bar{X}_{k,\cdot}$.
In order to test the non-inferiority or superiority hypothesis $H_{0}$, Wald-type statistics are often considered. 
Here, the Wald-type statistic $T_n$ is obtained by substituting the parameters $\mu_k$ in the hypothesis $H_0$ from Equation \eqref{Eq:H0_rearranged} by the means $\bar{X}_{k,\cdot},\, k=E,R,P$, and dividing the resulting term by an estimator of its standard deviation, i.e.
\begin{align}
T_n=T_n(\mathbf{X}_n)=\sqrt{n}\frac{\bar{X}_{E,\cdot}-\Delta\bar{X}_{R,\cdot}+(\Delta-1)\bar{X}_{P,\cdot}}{\hat{\sigma}}.
\end{align}
Here, the variance estimator $\hat{\sigma}^2$ is defined as 
\begin{align*}
\hat{\sigma}^{2} =
\frac{\hat{\sigma}^{2}_{E}}{w_E} + 
\Delta^2 \frac{\hat{\sigma}^{2}_{R}}{w_R} + 
(1-\Delta)^2 \frac{\hat{\sigma}^{2}_{P}}{w_P}
\end{align*}
with the group specific sample variances
\begin{align*}
\hat{\sigma}^{2}_{k} = \frac{1}{n_k - 1} \sum_{i=1}^{n_k}(X_{k,i}-\bar{X}_{k, \cdot})^2,\quad k=E,R,P.
\end{align*}
In the following, we introduce the studentized permutation test which uses the Wald-type statistic $T_n$ as a test statistic.
Let $(\tau(i))_{i\leq n}$ denote a random variable which is uniformly distributed on the group $S_n$ of all permutations of the first $n$ natural numbers. 
We denote the associated probability measure with $\tilde{\mathbb{P}}$. 
The probability measures $\tilde{\mathbb{P}}$ and $\mathbb{P}$ are independent.
We will use the notation $\tau_n(\mathbf{X}_n)=(X_{n, \tau(1)},\ldots,X_{n, \tau(n)})$ for the randomly permutated vector $\mathbf{X}_n$. 
For a given vector $\mathbf{X}_n$, the permutation statistic refers to the test statistic calculated with the permuted vector $\tau_{n}(\mathbf{X}_n)$. 
That is, the permutation statistic is the result obtained by the  mapping 
\begin{align*}
(\tau(i))_{i\leq n} \mapsto 
T_n\left(X_{n, \tau(1)},\ldots,X_{n, \tau(n)}\right)|\mathbf{X}_n.
\end{align*}
Then, for a given significance level $\alpha\in(0,1)$ the studentized permutation test $\varphi_{n}^{Perm}$ is the function
\begin{align*}
\varphi^{Perm}_{n}(\mathbf{X}_n)=
\begin{cases}
1 & T_n(\mathbf{X}_n)< c_{n}(\alpha) \\
0 & T_n(\mathbf{X}_n) \geq c_{n}(\alpha)
\end{cases}
\end{align*}
with $c_{n}(\alpha)$ the $\alpha$-quantile of the permutation distribution which is the largest number such that the inequality  
\begin{align*}
\tilde{\mathbb{P}}\Big(T_n\big(\tau_n(\mathbf{x}_{n})\big)<c_{n}(\alpha)\Big) \leq \alpha
\end{align*}
holds.
The studentized permutation test $\varphi_{n}^{Perm}$ is an asymptotically exact test. 
In other words, at the boundary of the null hypothesis which is given by the equation $(\mu_{P}-\mu_{E})=\Delta(\mu_{P}-\mu_{R})$, the expected value of the test function $\varphi_{n}^{Perm}$  with respect to the probability measure $\mathbb{P}$ converges to $\alpha$, that is
\begin{align}\label{eqn:permExact}
\lim_{n\to\infty}\mathbb{E}_{\mathbb{P}}\left[\varphi_{n}^{Perm}(\mathbf{X}_n)\right]=
\alpha.
\end{align}
Equation \eqref{eqn:permExact} follows immediately from the convergence of the distribution of the statistic $T_n$ against a normal distribution and the asymptotic normality of the permutation statistic. 
Asymptotic normality of the permutation statistic refers to the limit
\begin{align}\label{eqn:asymNormPerm}
\sup_{t\in \mathbb{R}}\left(\left\vert\tilde{\mathbb{P}}\big(T_n\left(\tau_n\left(\mathbf{X}_{n}\right)\right)<t\big\vert \mathbf{X}_{n} \big) - \Phi(t)\right\vert\right)\xrightarrow[n\to\infty]{\mathbb{P}}0.
\end{align}
Put into words, the asymptotic normality of the permutation statistic is the uniform convergence  in $\mathbb{P}$-probability of the cumulative distribution function of the permutation statistic against the cumulative distribution function $\Phi$ of a standard normal distribution. 
From the limit in Formula \eqref{eqn:asymNormPerm} it follows directly that the critical value $c_n(\alpha)$ of the permutation test converges to the $\alpha$-quantile of a standard normal distribution.
The asymptotic normality of the permutation statistic will be proven in the appendix  using the central limit theorem for conditional permutation distributions \cite{janssen1997studentized}. 
\section{Simulation study}
The small sample size properties of the studentized permutation test are studied in this section by means of a Monte-Carlo simulation study. 

For comparisons two Wald-type tests, which use the same Wald-type statistic $T_n$ as the studentized permutation test, are included into the simulation study. 
The first Wald-type test rejects the  hypothesis $H_0$  if the test statistic $T_n $ is smaller than  the $\alpha$-quantile $q_\alpha$ of a standard normal distribution. 
This test can be written as the function 
\begin{align*}
\phi_{n}^{WTn}(\mathbf{X}_n) = 
\begin{cases}
1 & T_n(\mathbf{X}_n)< q_{\alpha} \\
0 & T_n(\mathbf{X}_n) \geq q_{\alpha}
\end{cases}.
\end{align*}
However, the Wald-type test $\phi_{n}^{WTn}$ in general only meets the target significance level for large sample sizes. 
For small sample sizes, the distribution of the test statistic $T_n$ can be approximated by a t-distribution with degrees of freedom calculated by the Welch approximation \cite{welch1938significance} as done by Hasler \textit{et al.}  \cite{hasler2008assessing}, that is 
\begin{align*}
\hat{\nu} = 
\frac{\left(\frac{1}{n_{E}}\hat{\sigma}^2_{E} +\frac{\Delta^2}{n_{R}}\hat{\sigma}^2_{R} + \frac{(1-\Delta)^2}{n_{P}}\hat{\sigma}^2_{P}\right)^2}{\frac{1}{n^2_{E}(n_{E}-1)}\hat{\sigma}^4_{E} +\frac{\Delta^4}{n^2_{R}(n_{R}-1)}\hat{\sigma}^4_{R} + \frac{(1-\Delta)^4}{n^2_{P}(n_{P}-1)}\hat{\sigma}^4_{P}}.
\end{align*}
Thus, we define the Wald-type test which rejects the null hypothesis $H_0$ when the test statistic $T_n$ is smaller than the $\alpha$-quantile of a t-distribution with degrees of freedom $\hat{\nu}$ , i.e.
\begin{align*}
\phi_{n}^{WTt}(\mathbf{X}_n) = 
\begin{cases}
1 & T_n(\mathbf{X}_n)< t_{\alpha, \hat{\nu}} \\
0 & T_n(\mathbf{X}_n) \geq t_{\alpha, \hat{\nu}}
\end{cases}.
\end{align*}
Both Wald-type tests $\phi_{n}^{WTn}$ and $\phi_{n}^{WTt}$ are included in the simulation study for the sake of comparison.
The performance characteristic of interest in the Monte-Carlo simulation study is whether the tests meet the target significance level $\alpha=0.025$.
The tests' performance is studied for continuous data in the first part and for count data in the second part. 
The continuous data is generated using a normal, lognormal, and Chi-squared distribution, respectively, and the focus will be on the effect of skewness and unequal variances on the performance of the tests.  
For count data, the Poisson and negative binomial distribution are considered as motivated by the distribution of lesions counts in clinical trials in multiple sclerosis.
Moreover, the simulation study focuses on the non-inferiority hypothesis for small sample sizes. 
Each simulated type I error is the result of 25\,000 replications which corresponds to a Monte-Carlo error of less than $0.001$ for the significance level $\alpha=0.025$.
The rejection area of the studentized permutation test is calculated from 15\,000 permutations.
The simulations are conducted using the R computing environment (Version 3.1.2) \cite{Renvironment}. Both the permutation test and the Wald-type tests are implemented in the R-package \texttt{ThreeArmedTrials} which is available on CRAN. 
\subsection{Continuous data}
\begin{figure}[ht]
\centering
\includegraphics[width=1\textwidth]{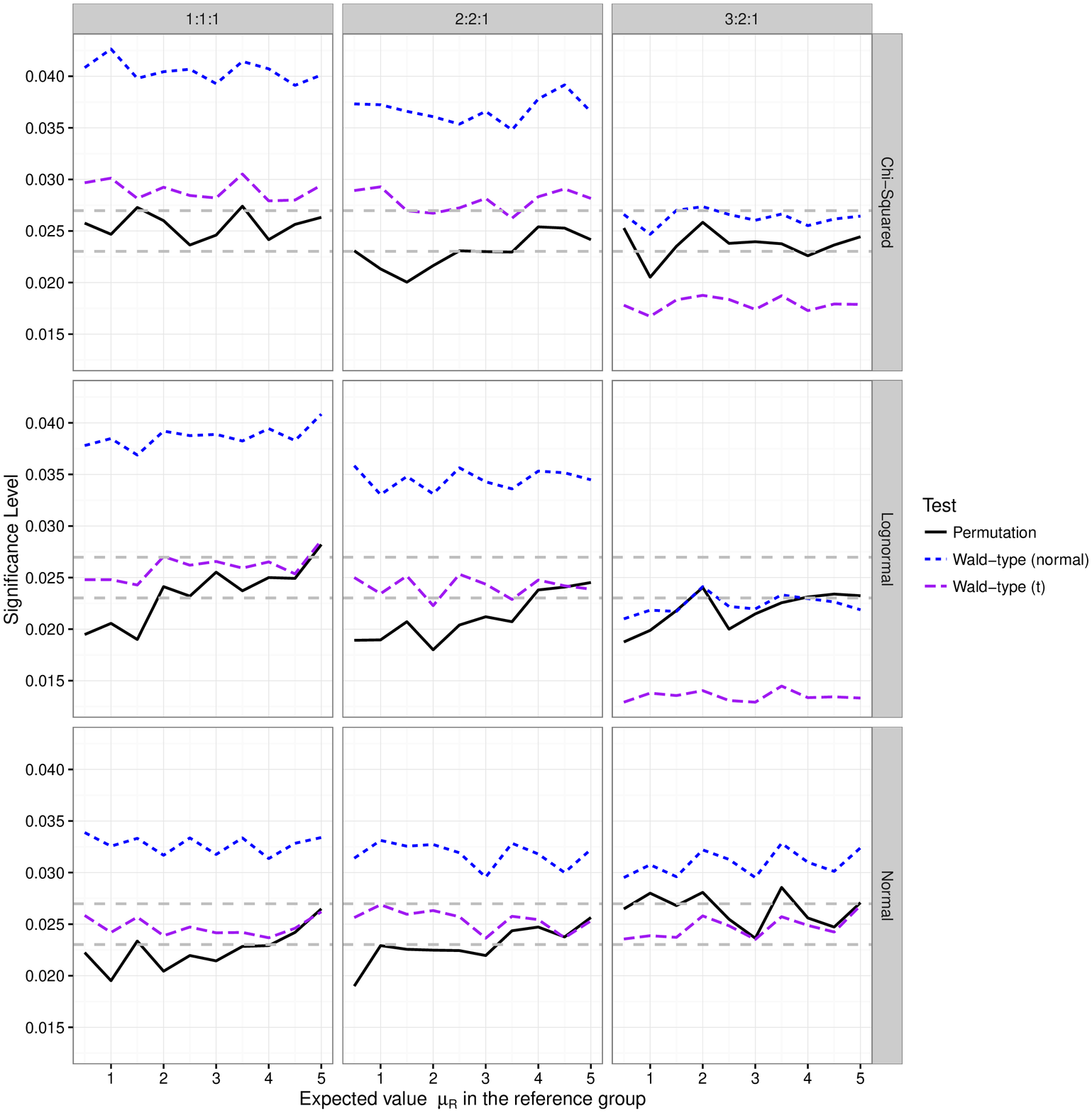}
\caption{Level of significance $\hat{\alpha}$ of the Wald-type tests and the permutation test for continuous data with equal group variances, $\sigma^2_E= \sigma^2_R= \sigma^2_P=1$. 
The dashed grey lines depict the area of $\alpha=0.025$ plus/minus two times the Monte-Carlo error. }
\label{Fig:ContinuousVar111}
\end{figure}

\begin{figure}[ht]
\centering
\includegraphics[width=1\textwidth]{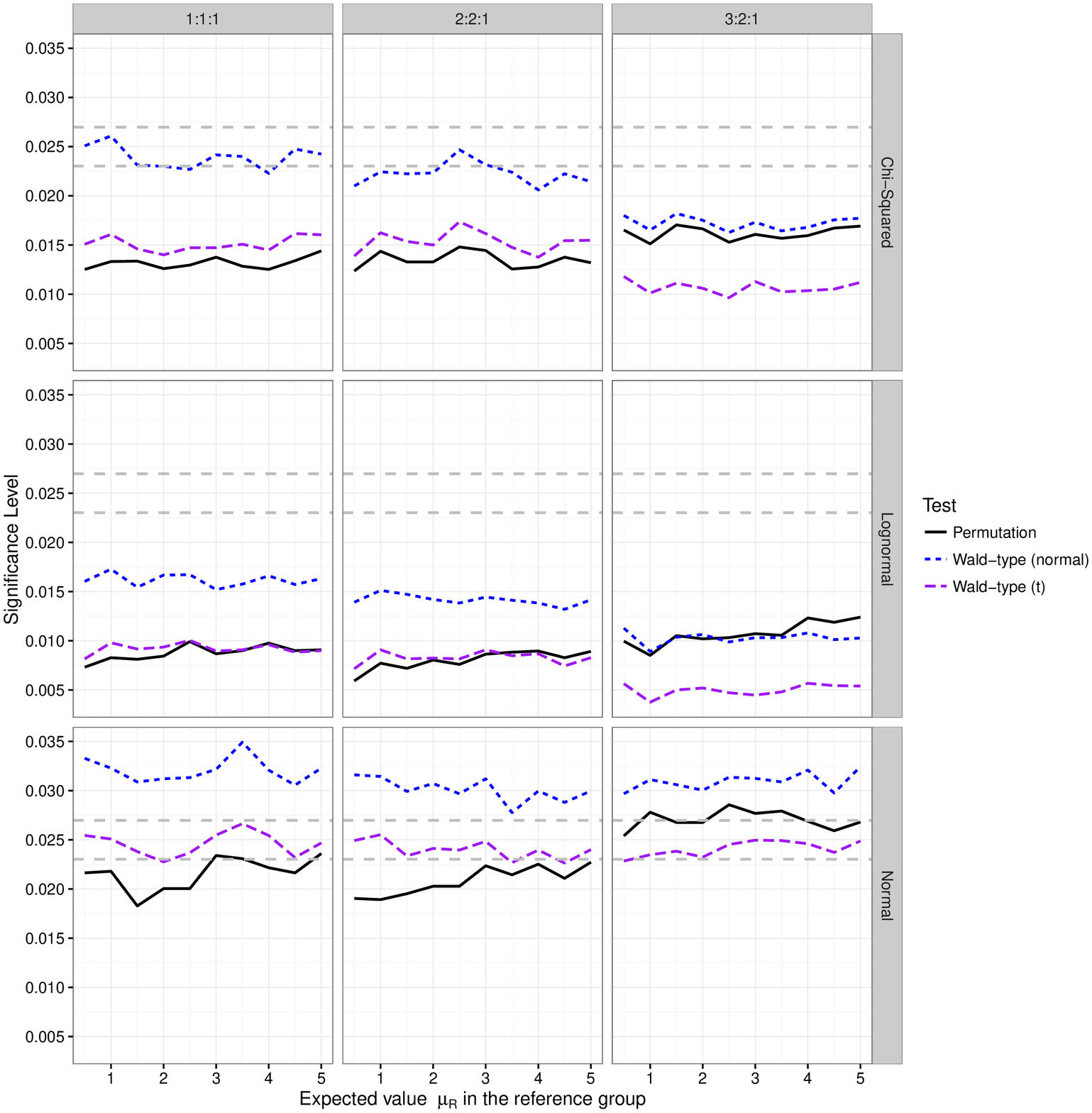}
\caption{Level of significance $\hat{\alpha}$ of the Wald-type tests and the permutation test for continuous data with unequal group variances $\sigma^2_E= 1,\, \sigma^2_R=2,\, \sigma^2_P=3$. 
The dashed grey lines depict the area of $\alpha=0.025$ plus/minus two times the Monte-Carlo error. }
\label{Fig:ContinuousVar123}
\end{figure}

In Table \ref{Table:ScenariosCont} the scenarios for the Monte-Carlo simulation study of the non-inferiority hypothesis tests are listed. 
\begin{table}[ht]
\begin{center}
\caption{Scenarios considered in the Monte-Carlo simulation study with continuous data.}
\label{Table:ScenariosCont}
\begin{tabular}{l*{1} {c}}
\toprule
\textbf{Parameter} & \textbf{Values}  \\ \midrule
One-sided significance level $\alpha$  & $0.025$\\
Clinical relevance margin $\Delta$ & 0.8\\
Mean $\mu_{P}$ & 5.5 \\
Mean $\mu_{R}$ & $0.5, 1, \ldots, 5$ \\
Group variances $(\sigma^2_E, \sigma^2_R, \sigma^2_P)$ & (1,1,1), (1,2,3) \\
Total sample size $n$  & $30$\\
Sample size allocations $n_{E}:n_{R}:n_{P}$  & 1:1:1, 2:2:1, 3:2:1\\
Distributions & Normal, Lognormal, Chi-squared\\
\bottomrule
\end{tabular}
\end{center}
\end{table}
The mean $\mu_{E}$ in the experimental treatment group is given by $\mu_E=\Delta \mu_R + (1-\Delta)\mu_P$. 
The lognormal distribution has location parameter $\mu=0$ and scale parameter $\sigma=1$. 
The Chi-square distribution has two degrees of freedom. 
For this simulation study, an observation $X_{k,i}$ is referred to as lognormal or chi-squared distributed when it is generated from a standardized lognormal or chi-squared distributed random.
More precisely, let  $X$ be a lognormal or Chi-squared distributed random variable, then the $X_{k,i}$ is calculated  by 
\begin{align*}
X_{k,i} = \left(\frac{X-\mathbb{E}[X]}{\sqrt{\operatorname{Var}(X)}}\right)\sigma_{k} + \mu_{k},\quad k = E,R,P.
\end{align*}
In Figures \ref{Fig:ContinuousVar111} and \ref{Fig:ContinuousVar123} the significance level of the different hypothesis tests is plotted against the reference group mean $\mu_R$ for the different sample size allocations and distributions.
Each frame in the 3x3-grids shows the results for one combination of sample size allocation and distribution.
In Figure \ref{Fig:ContinuousVar111} the group variances are identical and in Figure \ref{Fig:ContinuousVar123} the variances are unequal, that is $\sigma^2_E=1$, $\sigma^2_R=2$, and $\sigma^2_P=3$.
\\
Before we compare the performance of the different tests, it is worth noting that Figures \ref{Fig:ContinuousVar111} and \ref{Fig:ContinuousVar123} show that the Wald-type test $\phi_{n}^{WTt}$ meets the target significance level for normally distributed data. 
This is expected since the test $\phi_{n}^{WTt}$ is the respective parametric test for normal data \cite{hasler2008assessing}.
In practice, however, the distribution is unknown and to draw conclusions about the general performance of the tests, the comparison must be made for one sample size allocation across multiple distributions and variance structures. 
We begin with the balanced design.
The permutation test is on the edge of being conservative for equal variances and conservative for unequal variances. 
The Wald-type test $\phi_{n}^{WTt}$ tends to be liberal for some scenarios but mostly meets the target significance level for equal variances.
In contrast, the test $\phi_{n}^{WTt}$ is conservative when the variances are unequal and the data is lognormally or chi-squared distributed.
The permutation test throughout controls the significance level but is more conservative than the Wald-type test $\phi_{n}^{WTt}$.
The Wald-type test $\phi_{n}^{WTn}$ is either liberal, conservative, or meets the target significance level depending on the variance structure and the distribution.
The results for the sample size allocation 2:2:1 are qualitatively the same as for the balanced design.
For the sample size allocation 3:2:1, the permutation test meets the target significance level for equal variances but becomes conservative for unequal variances when the distribution is skewed. 
Here, the significance level of the permutation test is closer to the target than the significance level of the Wald-type test $\phi_{n}^{WTt}$ which is conservative, except for normally distributed data.
As for the other allocations, the significance level of the Wald-type test $\phi_{n}^{WTt}$ can be both deflated and inflated depending on the scenario.\\ \indent
Concluding, for all considered sample size allocations, the Wald-type test $\phi_{n}^{WTt}$ with a t-quantile as the critical value and the permutation test perform similar with respect to meeting the target significance level.
For the few scenarios where the  Wald-type test $\phi_{n}^{WTt}$ tends to be liberal, the permutation test controls the significance level.
The permutation test is the more conservative choice for a balanced design and an unbalanced design with sample size allocation 2:2:1. 
For a design with sample size allocation 3:2:1, the Wald-type test $\phi_{n}^{WTt}$ is more conservative than the permutation test. 
Eventually, both test can be recommended for continuous data. 
The Wald-type test $\phi_{n}^{WTn}$ with a normal quantile as the critical value should not be considered in practice.
\subsection{Count data}
\begin{figure}[ht]
\centering
\includegraphics[width=1\textwidth]{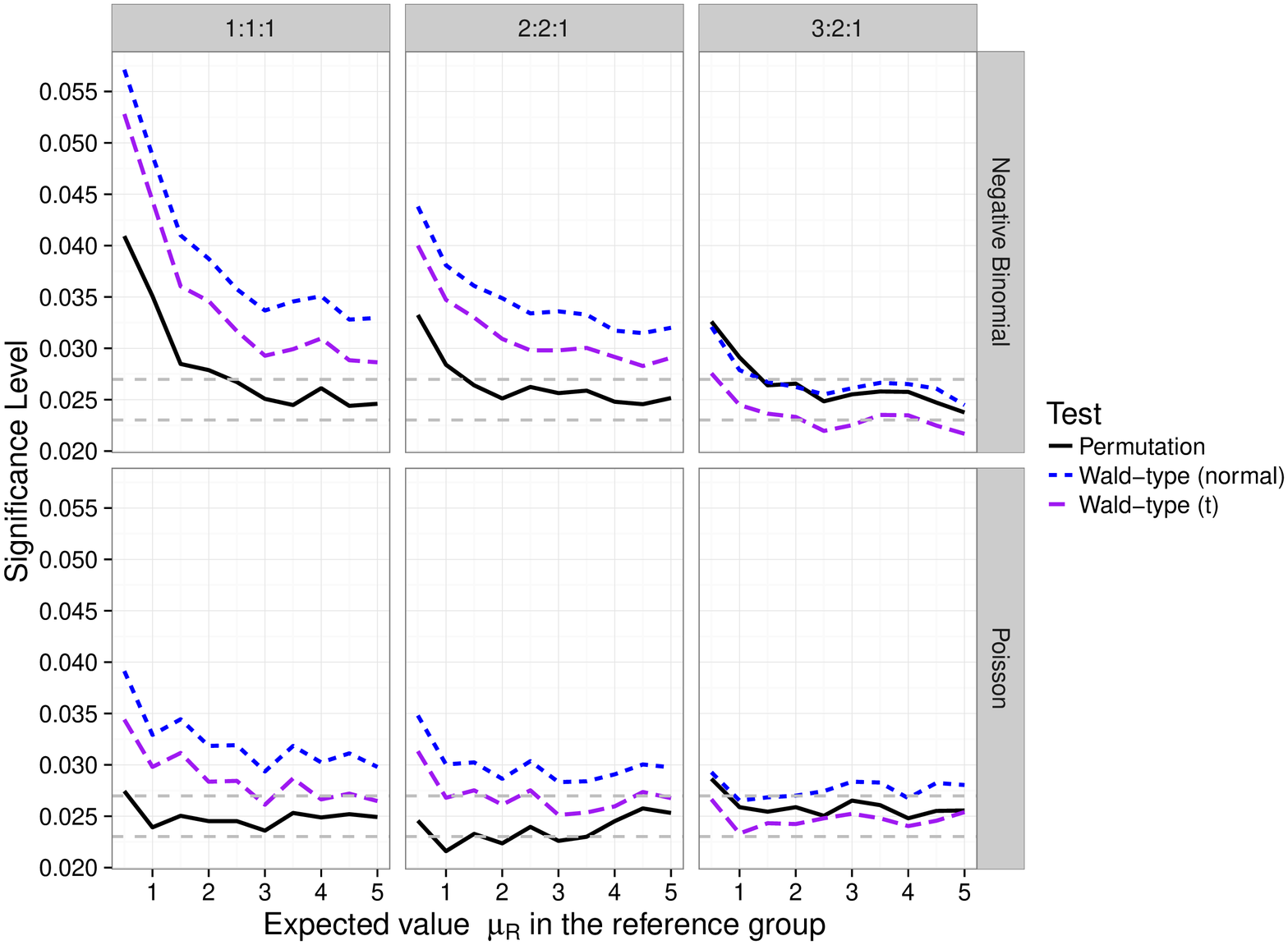}
\caption{Significance  level of the Wald-type tests and the permutation test for Poisson data and negative binomial data. 
The dashed grey lines depict the area of $\alpha=0.025$ plus/minus two times the Monte-Carlo error. }
\label{Fig:Count}
\end{figure}

\begin{figure}[ht]
\centering
\includegraphics[width=0.7\textwidth]{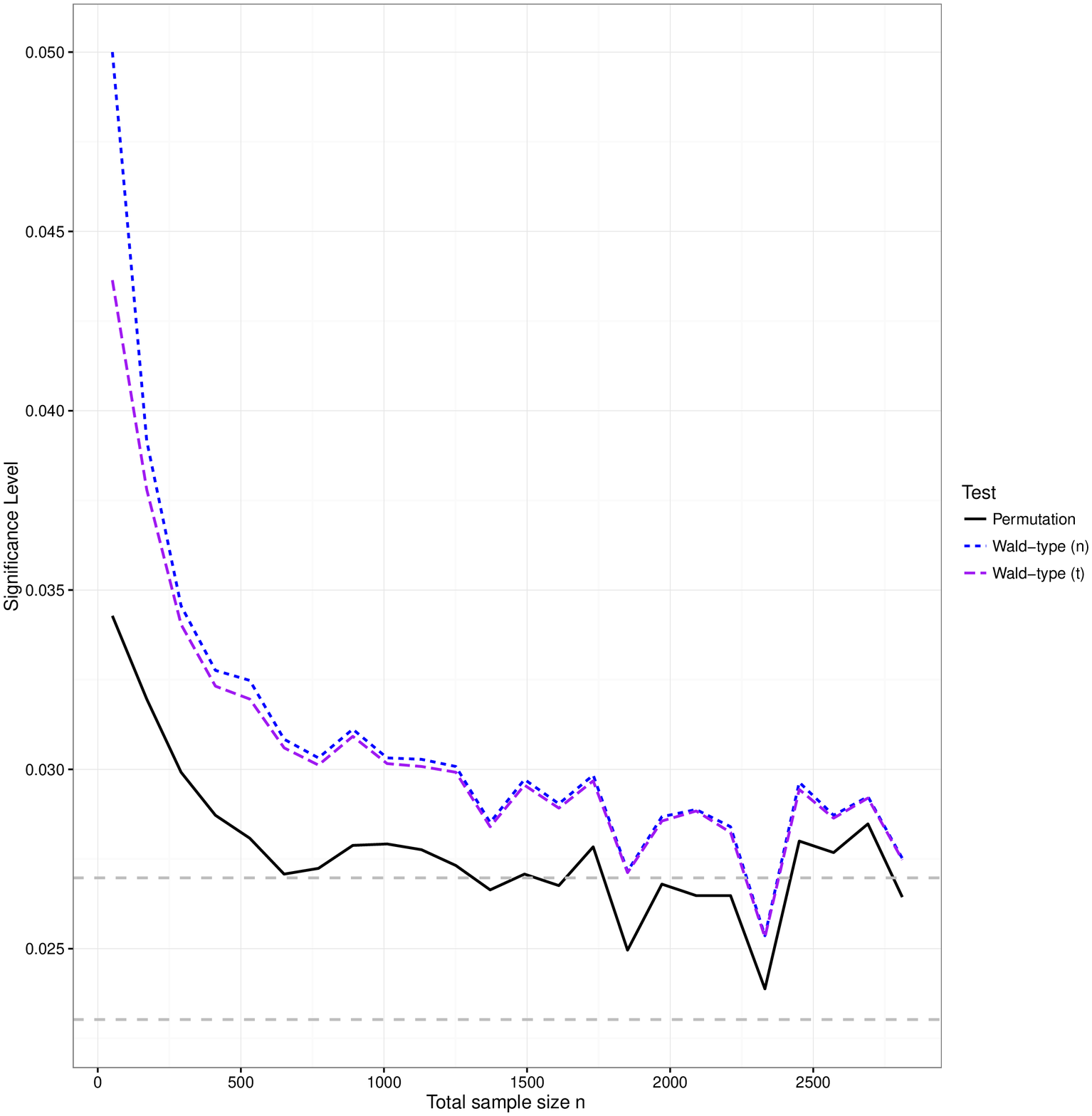}
\caption{Level of significance of the studentized permutation test and the two Wald-type tests against the total sample size $n$ for the scenario $(\mu_E, \mu_R, \mu_P)=(1.9, 1, 5.5)$ with $\kappa=3$ and sample size allocations 1:1:1. The dashed grey lines depict the target significance level $\alpha=0.025$ plus/minus two times the Monte-Carlo error.}
\label{Fig:SigLevelVsN}
\end{figure}

In this subsection we present the results of the Monte-Carlo simulation study for count data. 
The scenario choices in this subsection were motivated by the example about lesion counts in clinical trials in multiple sclerosis presented in Section 2.
Throughout this section, the expected value, which we will refer to as rate, in the placebo group is chosen to be $\mu_P=5.5$. 
Then, the rate in the reference treatment group is chosen to be between $\mu_R=0.5$ and $\mu_R=5$. 
The rate $\mu_E$ in the experimental treatment group is calculated by $\mu_E=\Delta \mu_R + (1-\Delta)\mu_P$ for the margin $\Delta=0.8$.
The group variance is defined as a multiple of the group rate, that is $\sigma^2_k=\kappa \mu_k$ with $\kappa=1,3$ for $k=E,R,P$. 
For $\kappa=1$, the random numbers are generated using a Poisson distribution. 
For $\kappa>1$, that is in the case of overdispersion, the random numbers are generated using a negative binomial distribution which is a two-parameter distribution with location parameter $\lambda>0$ and shape parameter $\phi > 0$. 
The expected value and variance are equal to $\lambda$ and $\lambda(1+\lambda\phi)$, respectively. 
Thus, for a given $\kappa$, the shape parameter is $\phi=(\kappa-1)/\lambda$. 
The negative binomial distribution converges in distribution to a Poisson distribution when $\phi$ approaches zero.
We consider the four sample size allocations $n_{E}:n_{R}:n_{P}=1\text{:}1\text{:}1, 2\text{:}2\text{:}1, 3\text{:}2\text{:}1$ and a total sample size of 60.
The example trial discussed in Section 2 had a larger sample size, however, we choose a trial size of 60 to also study the tests' performances for sample sizes of smaller phase II studies.
The scenarios are summarized in Table \ref{Table:ScenariosCount}. 
\begin{table}[ht]
\begin{center}
\caption{Scenarios considered in the Monte-Carlo simulation study with count data.}
\label{Table:ScenariosCount}
\begin{tabular}{l*{1} {c}}
\toprule
\textbf{Parameter} & \textbf{Values}  \\ \midrule
One-sided significance level $\alpha$  & $0.025$\\
Clinical relevance margin $\Delta$ &  0.8\\
Placebo rate $\mu_{P}$ & 5.5\\
Reference rate $\mu_{R}$ & $0.5, 1, \ldots, 5$ \\
Group variance $\sigma^2_k=\kappa \mu_k$ & $\kappa=1, 3$ \\
Total sample size $n$  & $60$\\
Sample size allocation $n_{E}:n_{R}:n_{P}$  & 1:1:1, 2:2:1, 3:2:1\\
\bottomrule
\end{tabular}
\end{center}
\end{table}
The results of the Monte-Carlo simulation study are shown in Figure \ref{Fig:Count}.
Figure \ref{Fig:Count} shows that the Wald-type test $\phi_{n}^{WTn}$ is liberal for all combinations of sample size allocation and distributions. 
The Wald-type test $\phi_{n}^{WTt}$ is liberal too, except for the allocation 3:2:1. 
The permutation tests meets the target significance level for Poisson data. 
However, the permutation test is liberal for negative binomial distributed data when the rate $\mu_R$ in the reference group is small. 
In these cases, the permutation test is less liberal than the Wald-type test competitors. 
For negative binomial distributed data we observe that all tests become less liberal as the rate $\mu_R$ increases.
Next, we study the convergence of the significance level of the studentized permutation test and the Wald-type test against the target significance level for increasing $n$. 
Figure \ref{Fig:SigLevelVsN} plots for scenario $(\mu_E, \mu_R, \mu_P)=(1.9, 1, 5.5)$ the significance level of the two tests against the total sample size for the sample size allocation 1:1:1. 
Figure \ref{Fig:SigLevelVsN} shows the convergence of the significance level  against the target for an increasing sample size $n$ for a scenario in which the tests do not meet the target significance level $\alpha=0.025$ for small sample sizes. 
This graphic highlights that for the tests a considerable increase of the sample size is required to reach the desired significance level.\\ \indent
In conclusion, the studentized permutation test should be preferred over the Wald-type test alternatives when testing the retention of effect hypothesis with (overdispersed) count data. 
Even though the permutation test is liberal in a few scenarios, it is less liberal than the Wald-type tests in the respective scenarios. 
\section{Discussion}
In this publication, we proposed a studentized permutation test for testing non-inferiority and superiority in three-arm trials in the `gold standard' design, that is trials with an experimental treatment, an active control, and a placebo. 
The studentized permutation test is an asymptotic test which does not require any distributional assumptions on the data, but only a finite expected value and variance. 
We compared the studentized permutation test in a Monte-Carlo simulation study with two Wald-type tests for normally, lognormally, chi-squared, Poisson, and negative binomial distributed data with particular emphasis on non-inferiority testing and small sample sizes. 
For continuous data, the Wald-type test with a normal quantile as the critical value is not recommended for application since it is either conservative or liberal. 
The studentized permutation test and the Wald-type test with a t-quantile as the critical value have overall a similar performance for continuous data.
For skewed data with unequal group variances both tests become conservative.
For use in practice both the studentized permutation test and the Wald-type test with a t-quantile can be recommended. Depending on the sample size allocation either one can be the more conservative choice.
For count data, the studentized permutation test outperforms the Wald-type test alternatives with respect to meeting the target significance level.
When the count data is overdispersed and the difference between the reference group mean and the placebo mean becomes larges, all considered test are liberal with the permutation test being the least liberal among the studied tests.
In practice, the studentized permutation test is recommended to analyze three-arm trials with count data when the sample sizes are small or no distributional assumptions can be made. 
This is an important finding because count endpoints are common in clinical trials.
While this work was motivated by the number of lesions in multiple sclerosis, other count endpoints include the number of hospitalizations in heart failure \cite{rogers2014analysing} and the number of exacerbations in COPD \cite{keene2008statistical}. \\ \indent
The adjustment for covariates is used in clinical trials to account for factors which have an influence on the primary endpoint. Covariates are briefly addressed in ICH guideline E9 \cite{ICHE9} and discussed in detail by the Committee for Proprietary Medicinal Products (CPMP) \cite{CPMP}.
The statistical model considered in this publication does not allow the adjustments for covariates. 
Especially under consideration of recently published results for studentized permutation tests in general factorial designs \cite{pauly2015asymptotic}, future work could focus on the extension of the proposed studentized permutation test to models including covariates.\\ \indent
In this publication non-inferiority of the experimental treatment compared to the reference has been defined with the retention of effect hypothesis which incorporates the placebo response into the definition of non-inferiority.
Hida and Tango \cite{hida2011three} and Stucke and Kieser \cite{stucke2012general} proposed a pairwise comparison of the experimental treatment and the reference to assess non-inferiority in a three-arm trial based on absolute margins. 
This pairwise assessment of non-inferiority is then part of a global hypothesis which additionally assesses superiority of the active treatments to placebo. 
Future research could focus on applying studentized permutation tests to the respective global hypothesis in three-arm trials with non-inferiority defined by pairwise comparison. 
\appendix
\section{Asymptotic normality of the permutation statistic}
We will now prove the asymptotic normality of the permutation statistic, that is
\begin{align*}
\sup_{t\in \mathbb{R}}\left(\left\vert\tilde{\mathbb{P}}\left(T_n\left(\tau\left(\mathbf{X}_{n}\right)\right)<t\big\vert \mathbf{X}_{n} \right) - \Phi(t)\right\vert\right)\xrightarrow[n\to\infty]{\mathbb{P}}0,
\end{align*}
by showing that the random vector $\mathbf{X}_n$ and the corresponding permutation statistic as defined in Section 4 fulfil the conditions of the central limit theorem for conditional permutation distributions \cite[Theorem 3.3]{janssen1997studentized}.
As preparation for the proof, we note that the test statistic $T_n(\mathbf{X}_n)$ can be written as the linear statistic $\sum_{i=1}^{n}c_{n,i}X_{n,i}$ with 
\begin{align*}
c_{n,i}:=
\sqrt{\dfrac{n_{E}n_{R}n_{P}}{n_{R}n_{P} + \Delta^2 n_{E}n_{P} + (\Delta-1)^2 n_{E}n_{R}}} \times 
\begin{cases}
-\frac{1}{n_{E}} & i=1,\ldots, n_{E}\\
\frac{\Delta}{n_{R}} & i=n_{E}+1,\ldots, n_{E}+n_{R}\\
\frac{1-\Delta}{n_{P}} & i=n_{E}+n_{R}+1,\ldots, n
\end{cases}
\end{align*}
a scheme of  regression coefficients for each $n\in \mathbb{N}$.
Then, according to the central limit theorem for conditional permutation distributions, the permutation statistic is asymptotically normally distributed if the following five conditions are fulfilled.
\begin{enumerate}
\item For each $n\in \mathbb{N}$ the sum of the squared regression coefficients and the sum of the regression coefficients is equal to one and zero, respectively:
\begin{align*}
\sum_{i=1}^{n}c_{n,i}^{2} & = 1\quad \forall \, n\in \mathbb{N} ,\\
\sum_{i=1}^{n}c_{n,i} & = 0\quad \forall \, n\in \mathbb{N} .
\end{align*}
Moreover, the maximum of the sequence $(c_{n,i})_{i\leq n}$ converges to zero:
\begin{align*}
\max_{1\leq i \leq n} |c_{n,i}| \xrightarrow{n\to \infty} 0.
\end{align*}
\item It holds that 
\begin{align*}
\liminf_{n\to \infty} \frac{1}{n} \sum_{i=1}^{n} (X_{n,i} - \bar{X}_{n,\cdot})^2 > 0 \qquad \mathbb{P} - a.s.
\end{align*}
\item There exists $\tilde{\sigma}>0$ such that
\begin{align*}
\dfrac{1}{\hat{\sigma}^2_{Perm}(\tau_{n}(\mathbf{X}_{n}))} 
\dfrac{1}{n}\sum_{i=1}^{n}(X_{n,i} - \bar{X}_{n,\cdot})^2
\xrightarrow[n\to \infty]{\mathbb{P} \times \tilde{\mathbb{P}} } \tilde{\sigma}^2.
\end{align*}
\item For $d\to \infty$ it holds:
\begin{align*}
\limsup_{n\to \infty} \dfrac{1}{n}\sum_{i=1}^{n}(X_{n,i} - \bar{X}_{n,\cdot})^2\mathbf{1}_{[d,\infty)} \left( |X_{n,i} - \bar{X}_{n,\cdot}|\right) \to 0 \qquad \mathbb{P}-\text{a.s.}
\end{align*}
\end{enumerate}
In the remainder of this section, we show that the studentized permutation test for the retention of effect hypothesis fulfills conditions 1. to 4. 
Thereto, without loss of generality, we assume that the expectation of the average of $\mathbf{X}_{n}$ is equal to zero,
\begin{align}
\label{eqn:AssumptionAsymNorm}
\mathbb{E}\left[ 
\frac{1}{n} \sum_{i=1}^{n}X_{n,i}
\right]
=0,
\end{align}
because the statistic $T_{n}(\mathbf{X_{n}})$ is invariant under the same shift for each $X_{n,i}$. Of course, the shift does not effect the variance of the random variables. 
\begin{enumerate}
\item 
The sums and the limit can be easily calculated which will not be shown here. 
\item 
With the assumption from Formula \eqref{eqn:AssumptionAsymNorm} that the expectation of the average $\bar{X}_{n,\cdot}$ is zero  and the strong law of large numbers, the average $\bar{X}_{n,\cdot}$ converges almost surely to zero. By means of the continuous mapping theorem, the squared average $\bar{X}_{n,\cdot}^2$ converges almost surely to zero.
With the property that the sum of three sequences of random variables converges almost surely if each of the sequences converges almost surely as well as with the strong law of large numbers, the average of the squared random variables converges almost surely:
\begin{align*}
\dfrac{1}{n}\sum_{i=1}^{n}X_{n,i}^2 \xrightarrow{n\to \infty} w_{E} (\sigma^2_{E} + \mu_{E}^2 ) + w_{R} (\sigma^2_{R} + \mu_{R}^2 ) + w_{P} (\sigma^2_{P} + \mu_{P}^2 ) \qquad \mathbb{P} \text{-a.s.}
\end{align*}
The assertion follows with the algebraic formula of the sample variance.
\item 
We define $\tilde{\sigma}:=1$ and prove the convergence 
\begin{align}\label{proof.sigma.tilde}
\dfrac{1}{\hat{\sigma}^2(\tau_{n}\left(\mathbf{X_{n}}\right))}\dfrac{1}{n}\sum_{i=1}^{n}(X_{n,i} - \bar{X}_{n,\cdot})^2\xrightarrow[\mathbb{P} \times \tilde{\mathbb{P}}]{n\to \infty }1.
\end{align}
Thereto, we show that the variance estimators $\hat{\sigma}^2(\tau_{n}\left(\mathbf{X_{n}}\right))$ and $\dfrac{1}{n}\sum_{i=1}^{n}(X_{n,i} - \bar{X}_{n,\cdot})^2$ converge in $\mathbb{P} \times \tilde{\mathbb{P}}$-probability to the same limit, that is
\begin{align}
 w_{E} (\sigma^2_{E} + \mu_{E}^2 ) + w_{R} (\sigma^2_{R} + \mu_{R}^2 ) + w_{P} (\sigma^2_{P} + \mu_{P}^2).
\label{Eq:Limit.Sigma2.Perm2}
\end{align}
The limit of $\dfrac{1}{n}\sum_{i=1}^{n}(X_{n,i} - \bar{X}_{n,\cdot})^2$ follows immediately from the second condition since we showed $\mathbb{P}$--a.s. convergence which implies $\mathbb{P}\times \tilde{\mathbb{P}}$--a.s. convergence which in turn implies the convergence in $\mathbb{P}\times \tilde{\mathbb{P}}$ -- probability.\\
To prove the convergence of the variance estimator $\hat{\sigma}^2(\tau_{n}\left(\mathbf{X_{n}}\right))$,  we decompose it by means of the algebraic formula for the sample variance, that is
\begin{align*}
\hat{\sigma}^2(\tau_{n}\left(\mathbf{X_{n}}\right)) = W_{n,1} - W_{n,2}^2 - W_{n,3}^2 - W_{n,4}^2
\end{align*}
with 
\begin{align*}
& W_{n,1} := \sum_{i=1}^{n}d_{n,i}X_{n,\tau(i)}^{2}, \qquad
 W_{n,2} := \dfrac{1}{\sqrt{n_E}}\sum_{i=1}^{n_E}\sqrt{d_{n,i}}X_{n,\tau(i)}, \\
& W_{n,3} := \dfrac{1}{\sqrt{n_R}}\sum_{i=n_E+1}^{n_E+n_R+1}\sqrt{d_{n,i}}X_{n,\tau(i)}, \qquad
 W_{n,4} := \dfrac{1}{\sqrt{n_P}}\sum_{i=n_E+n_R+1}^{n}\sqrt{d_{n,i}}X_{n,\tau(i)}.
\end{align*}
The sequence $(d_{n,i})_{i\leq n}$ is defined as
\begin{align*}
d_{n,i} := \dfrac{n_{E}n_{R}n_{P}}{n_{P}n_{R} + \Delta^2 n_{P}n_{R} + (\Delta - 1)^2 n_{E}n_{R}} \times 
\begin{cases}
\frac{1}{n_{E}(n_{E}-1)} & i=1,\ldots, n_{E}\\
\frac{\Delta ^2}{n_{R}(n_{R}-1)} & i=n_{E}+1,\ldots, n_{E}+n_{R}\\
\frac{(\Delta -1)^2}{n_{P}(n_{P}-1)} & i=n_{E}+n_{R}+1,\ldots, n
\end{cases}.
\end{align*}
Here, $X_{n,\tau(i)}$ denotes the i-th entry of the vector $\tau_{n}(\mathbf{X_{n}})$.
We prove the convergence of the variance estimator $\hat{\sigma}^2 \left( \tau_{n}\left(\mathbf{X_{n}}\right) \right)$ by showing that $W_{n,2}, W_{n,3}$, and $W_{n,4}$ converge in $\mathbb{P} \times \tilde{\mathbb{P}}$--probability to zero as well as that $W_{n,1}$ converges to the limit stated in \eqref{Eq:Limit.Sigma2.Perm2}. 
The proofs for the convergence of $W_{n,2}, W_{n,3}$, and $W_{n,4}$ are similar and, therefore, we only prove it for $W_{n,2}$. Said proof follows next. \\
Taking into account that $X_{n,\tau(i)}$ and $X_{n,\tau(1)}$ have the same distribution as well as that with probability $\frac{1}{n}$ the random variable $X_{n,\tau(1)}$ is equal to the random variable $X_{n,i},\, i=1,\ldots,n$, the expectation $\mathbb{E}_{\tilde{\mathbb{P}}}\left[  W_{n,2} \right]$ is equal to $\sqrt{\kappa}\, n_{E} \bar{X}_{n,\cdot}$.
Thus,  $\mathbb{E}_{\mathbb{P}\times \tilde{\mathbb{P}}}[W_{n,2}]=0$ follows immediately from the independence of $\mathbb{P}$ and $\tilde{\mathbb{P}}$ as well as the assumption that the average $\bar{X}_{n,\cdot}$ has expectation zero. 
Let $\varepsilon>0$ be an arbitrary real number, by applying Markov's inequality, we obtain 
\begin{align*}
 (\mathbb{P}\times \tilde{\mathbb{P}})\left( \left\vert W_{n,2} \right\vert \geq \varepsilon \right)
\leq \frac{1}{\varepsilon^2} \operatorname{Var}_{\mathbb{P}\times \tilde{\mathbb{P}}}\left[  W_{n,2}  \right].
\end{align*}
Due to the independence of $\mathbb{P}$ and $\tilde{\mathbb{P}}$ and the law of total variance, we obtain 
\begin{align*}
\operatorname{Var}_{\mathbb{P}\times \tilde{\mathbb{P}}}\left[  W_{n,2}  \right] = \mathbb{E}_{\mathbb{P}}\left[\operatorname{Var}_{ \tilde{\mathbb{P}}}\left[  W_{n,2} \right]\right]
+\operatorname{Var}_{\mathbb{P}}\left[\mathbb{E}_{ \tilde{\mathbb{P}}}\left[  W_{n,2} \right]\right].
\end{align*}
Hence, $W_{n,2}$ converges in probability to zero if the expectation $\mathbb{E}_{\mathbb{P}}\left[\operatorname{Var}_{ \tilde{\mathbb{P}}}\left[  W_{n,2} \right]\right]$ and the variance
 $\operatorname{Var}_{\mathbb{P}}\left[\mathbb{E}_{ \tilde{\mathbb{P}}}\left[  W_{n,2} \right]\right]$ converge to zero as $n$ tends to infinity.
 For the sake of readability, we define 
\begin{align*}
\kappa := \dfrac{n_{R}n_{P}}{(n_{R}n_{P} + \Delta^2 n_{E}n_{P} + (\Delta-1)^2 n_{E}n_{R})n_{E}(n_{E}-1)}.
\end{align*}
With $\kappa=\kappa(n)\in \mathcal{O}(\frac{1}{n^2})$, we obtain 
\begin{align}\label{Eq:Limit.Sigma2.Perm3}
\lim_{n\to \infty} \operatorname{Var}_{\mathbb{P}}\left[\mathbb{E}_{ \tilde{\mathbb{P}}}\left[  W_{n,2} \right]\right]
\leq  \lim_{n\to \infty}  \kappa \frac{n_{E}}{n}\max\limits_{1\leq i\leq n}\operatorname{Var}_{\mathbb{P}}[X_{n,i}] = 0.
\end{align}
Moreover, to prove that the expectation $\mathbb{E}_{\mathbb{P}}\left[\operatorname{Var}_{ \tilde{\mathbb{P}}}\left[  W_{n,2} \right]\right]$  converges to zero, we rearrange the variance $\operatorname{Var}_{ \tilde{\mathbb{P}}}\left[  W_{n,2} \right]$ to
\begin{align*}
\kappa\, \mathbb{E}_{ \tilde{\mathbb{P}}}\left[  \sum_{i, j=1,\atop i\neq j }^{n_{E}}X_{n,\tau(i)}X_{n,\tau(j)} + \sum_{i=1}^{n_{E}}X_{n,\tau(i)}^{2} \right] - \kappa\, n_{E}^2\bar{X}_{n,\cdot}^{2}.
\end{align*}
For $i\neq j$ and  $i'\neq j'$, the random variables $X_{n,\tau(i)}X_{n,\tau(j)}$ and  $X_{n,\tau(i')}X_{n,\tau(j')}$ are identically distributed with respect to $\tilde{\mathbb{P}}$ and with probability $1/(n(n-1))$ the random variable $X_{n,\tau(1)}X_{n,\tau(2)}$ is equal to $X_{n,i}X_{n,j}$. Thus, we obtain
\begin{align*}
 \operatorname{Var}_{ \tilde{\mathbb{P}}}\left[  W_{n,2} \right]
= & \kappa \frac{n_{E}(n-n_{E})}{n(n-1)}\sum_{i=1}^{n}(X_{n,i}-\bar{X}_{n,\cdot})^2.
\end{align*}
We already proved that the sample variance converges $\mathbb{P}$ almost surely and with $k=k(n)\in\mathcal{O}(1/n^2)$, it follows that the expectation $\mathbb{E}_{\mathbb{P}}[\operatorname{Var}_{ \tilde{\mathbb{P}}}\left[  W_{n,2} \right]]$ converges to zero as $n$ approaches infinity. Thus, $W_{n,2}$ and analogously $W_{n,3}$ as well as $W_{n,4}$ converge in $\mathbb{P}\times \tilde{\mathbb{P}}$-probability to zero.\\
To prove the convergence of the variance estimator $\hat{\sigma}^2_{Perm} \left( \tau_{n}(\mathbf{X_{n}}) \right)$, it remains to show that $W_{n,1}$ converges in probability to $W_{1}:=w_{E} (\sigma^2_{E} + \mu_{E}^2 ) + w_{R} (\sigma^2_{R} + \mu_{R}^2 ) + w_{P} (\sigma^2_{P} + \mu_{P}^2 )$. 
Due to the independence of $\mathbb{P}$ and $\tilde{\mathbb{P}}$, the expectation of $W_{n,1}$ with respect to $\mathbb{P}\times \tilde{\mathbb{P}}$ is given by
\begin{align*}
\mathbb{E}_{\mathbb{P}\times \tilde{\mathbb{P}}}[W_{n,1}]
= d_{n,\cdot} \frac{1}{n}\left( n_{E}(\sigma^{2}_{E} + \mu_{E}^{2}) +n_{R}(\sigma^{2}_{R} + \mu_{R}^{2}) + n_{P}(\sigma^{2}_{P} + \mu_{P}^{2}) \right).
\end{align*}
It follows that the expectation $\mathbb{E}_{\mathbb{P}\times \tilde{\mathbb{P}}}[W_{n,1}]$ converges to $W_{1}$.
Let $\varepsilon>0$ be an arbitrary real number and $n$ sufficiently large such that $|\mathbb{E}_{\mathbb{P} \times \tilde{\mathbb{P}}}[W_{n,1}]-W_{1}|<\varepsilon$, by applying Markov's inequality, we obtain
\begin{align*}
& (\mathbb{P}\times \tilde{\mathbb{P}})\left( \left\vert W_{n,1} -W_1\right\vert \geq \varepsilon \right)
\leq  \frac{1}{\left( \varepsilon -\big|\mathbb{E}_{\mathbb{P} \times \tilde{\mathbb{P}}}[W_{n,1}]-W_{1}\big|\right)^2} \operatorname{Var}_{\mathbb{P}\times \tilde{\mathbb{P}}}\left[  W_{n,1}  \right].
\end{align*}
To show that the right side converges to zero as $n$ approaches infinity, we note that for the sums  $d_{n,\cdot}:=\sum_{i=1}^{n}d_{n,i}$ and $\sum_{i=1}^{n}d^2_{n,i}$, we have the asymptotic properties 
$\lim_{n\to \infty} d_{n,\cdot}=1$ 
as well as 
$\lim_{n\to \infty} \sum_{i=1}^{n}d^2_{n,i}=0$.
For both limits, we took into account that none of the three groups vanish asymptotically, that is $\lim_{n\to \infty}n_{k}/n=w_{k}\in (0,1)$.
The variance of $W_{n,1}$ is equal to
\begin{align*}
\operatorname{Var}_{\mathbb{P}\times \tilde{\mathbb{P}}}[W_{n,1}] = \mathbb{E}_{\mathbb{P}}\left[\operatorname{Var}_{\tilde{\mathbb{P}}}[W_{n,1}]\right] + \operatorname{Var}_{\mathbb{P}}\left[\mathbb{E}_{\tilde{\mathbb{P}}}[W_{n,1}]\right].
\end{align*}
Since the forth moment of $X_{n,i}$ with $i=1,\ldots,n$ is bounded, for the second term follows
\begin{align*}
&\operatorname{Var}_{\mathbb{P}}\left[\mathbb{E}_{\tilde{\mathbb{P}}}[W_{n,1}]\right]= d_{n,\cdot}^{2} \operatorname{Var}_{\mathbb{P}}\left[\frac{1}{n}\sum_{i=1}^{n}X^{2}_{n,i}\right]
 \leq  d_{n,\cdot}^{2} \frac{1}{n} \max\limits_{1\leq i \leq n} \mathbb{E}_{\mathbb{P}}\left[X_{n,i}^{4}\right].
\end{align*}
Hence, the variance $\operatorname{Var}_{\mathbb{P}}\left[\mathbb{E}_{\tilde{\mathbb{P}}}[W_{n,1}]\right]$ converges to zero as $n$ approaches infinity.
It remains to prove that $\mathbb{E}_{\mathbb{P}}\left[\operatorname{Var}_{\tilde{\mathbb{P}}}[W_{n,1}]\right]$ converges to zero. 
Thereto, we calculate the variance $\operatorname{Var}_{\tilde{\mathbb{P}}}[W_{n,1}]$.
With $\bar{d}_{n,\cdot}:=d_{n,\cdot} / n$, we obtain the variance $\operatorname{Var}_{\tilde{\mathbb{P}}}[W_{n,1}] $ 
\begin{align*}
\operatorname{Var}_{\tilde{\mathbb{P}}}[W_{n,1}] 
= \sum_{i=1}^{n}(d_{n,i}-\bar{d}_{n,\cdot})^2 \frac{1}{n-1}\sum_{j=1}^{n}\left(X_{n,j}^{2} - \frac{1}{n}\sum_{i=1}^{n}X_{n,i}^{2}\right)^2.
\end{align*}
Since the forth moment of $X_{n,i}$ with $i=1,\ldots,n$ is bounded, the term 
\begin{align*}
\frac{1}{n-1}\sum_{j=1}^{n}\left(X_{n,j}^{2} - \frac{1}{n}\sum_{i=1}^{n}X_{n,i}^{2}\right)^2
\end{align*}
converges in $\mathbb{P}$-probability to a finite limit. Since $\sum_{i}(d_{n,i}-\bar{d}_{n,\cdot})^2$ converges to zero, it follows that the expectation $\mathbb{E}_{\mathbb{P}}[\operatorname{Var}_{\tilde{\mathbb{P}}}[W_{n,1}]]$ also converges to zero. Therefore, $W_{n,1}$ converges in  $\mathbb{P}\times \tilde{\mathbb{P}}$-probability to $W_{1}$. 
\item 
With $\limsup_{n\to \infty}(a_{n}+b_{n})\leq \limsup_{n\to \infty}a_{n}+\limsup_{n\to \infty}b_{n}$ and the algebraic formula for the variance, we obtain $\mathbb{P}$ almost surly the inequality
\begin{align*}
& \lim_{d\to \infty} \limsup_{n\to \infty} \dfrac{1}{n} \sum_{i=1}^{n} \left( X_{n,i} - \bar{X}_{n,\cdot} \right)^2 \mathbf{1}_{[d,\infty )}( |X_{n,i} - \bar{X}_{n,\cdot} | ) \\
\leq & \lim_{d\to \infty} \left( \limsup_{n\to \infty} \dfrac{1}{n} \sum_{i=1}^{n}  X_{n,i}^2 \mathbf{1}_{[d,\infty )}( |X_{n,i} - \bar{X}_{n,\cdot} | ) 
+ \limsup_{n\to \infty} \left( - \bar{X}^{2}_{n,\cdot}  \mathbf{1}_{[d,\infty )}( |X_{n,i} - \bar{X}_{n,\cdot} | )  \right) \right)
\end{align*}
Since $\bar{X}_{n,\cdot}$ converges $\mathbb{P}$ almost surly to zero, the second limes superior is zero for each $d$. Due to the strong law of large number which holds because the $\mathbb{E}[X_{n,i}^{2}]$ are bounded, for each $d$, the first limes superior is equal to 
\begin{align*}
 \limsup_{n\to \infty} \dfrac{1}{n} \sum_{i=1}^{n}  X_{n,i}^2 \mathbf{1}_{[d,\infty )}( |X_{n,i} - \bar{X}_{n,\cdot} | )=
\sum_{k=E,R,P} w_{k}  \mathbb{E}\left[ X_{k,1}^2 \mathbf{1}_{[d,\infty )}( |X_{k,1}  )\right].
\end{align*}
Since the expectation $\mathbb{E}[X_{k,1}^{2}]$ exists for $k=E,R,P$, the expectation $\mathbb{E}\left[ X_{k,1}^2 \mathbf{1}_{[d,\infty )}( |X_{k,1}|  )\right]$ converges to zero as $d$ approaches infinity.
\end{enumerate}

 \section*{Acknowledgement}
This research benefited from discussions on modeling recurrent event data at workshops supported by Deutsche Forschungsgemeinschaft under grant FR3070/1-1.
Tobias M\"utze is supported by the DZHK (German Centre for Cardiovascular Research).
Axel Munk acknowledges support of the Volkswagen Foundation within the Felix-Bernstein Institute for Mathematical Statistics in the Biosciences.



\bibliographystyle{model1-num-names}



\end{document}